\documentstyle[psfig,epsfig,12pt]{article}

\newcommand{\be}{\begin{equation}}
\newcommand{\ee}{\end{equation}}
\newcommand{\bea}{\begin{eqnarray}}
\newcommand{\eea}{\end{eqnarray}}

\title{Quantum Instantons and Quantum Chaos}

\author{
H.~Jirari$^{a}$, H.~Kr\"{o}ger$^{a}\footnote{Corresponding author, 
Email: hkroger@phy.ulaval.ca}$,
X.Q.~Luo$^{b,c}$, 
K.J.M.~Moriarty$^{d}$ 
and S.G.~Rubin$^{e}$ \\ [2mm]
{\small\sl $^{a}$D\'{e}partement de Physique, Universit\'{e} Laval, Qu\'{e}bec, Qu\'{e}bec G1K 7P4, Canada} \\ 
{\small\sl $^{b}$CCAST (World Laboratory), P.O. Box 8730, Bejing 100080, China} \\ 
{\small\sl $^{c}$Departement of Physics, Zhongshan University, Guangzhou 510275, China} \\
{\small\sl $^{d}$Department of Mathematics, Statistics and Computer Science,} \\
{\small\sl Dalhousie University, Halifax, N.S. B3H 3J5, Canada} \\
{\small\sl $^{e}$Moscow Engineering Physics Institute, Center for Cosmo-Particle Physics "Cosmion", Moscow, Russia} \\ 
}

\begin{document} 
\maketitle

\begin{abstract}
\noindent We suggest a closed form expression for the path integral 
of quantum transition amplitudes to construct a quantum action. 
Based on this we propose rigorous definitions of both, quantum instantons and quantum chaos. As an example we compute the quantum instanton of the double well potential. 
\end{abstract}

\setcounter{page}{0}

\newpage
\noindent {\bf 1. Introduction} \\
Instantons and chaos both play an important role in modern science.
Chaos occurs in physics, chemistry, biology, physiology, meteorology, 
economy etc. Both concepts are defined by classical physics. On the other hand, for instanton solutions occuring in quantum field theory of gauge theories, quantum effects are very important.
So far quantum effects have been computed mostly perturbatively (WKB).
The chaotic behavior of quantum liquids and quantum dots 
requires a quantum mechanical description. 
However, the concept of quantum chaos has eluded yet a rigorous definition, 
albeit a quantitative computation. In this work we propose how to define 
quantum instantons and quantum chaos. As an example, we compute 
quantitatively the quantum instanton for a 1-dim quantum system.

Chaotic phenomena were found in a number of quantum systems. For example, the hydrogen atom in a strong magnetic field shows strong irregularities in the spectrum \cite{Friedrich:89}. Irregular patterns were observed in the wave functions of the quantum mechanical model of the stadium billard \cite{McDonald:79}. Billard like boundary counditions have been realized experimentally in mesoscopic quantum systems, like quantum dots and quantum corrals, formed by atoms in semi-conductors \cite{Stockmann:99}.
While classical chaos theory, using Lyapunov exponents 
and Poincar\'e sections, is based on identifying trajectories in phase space, this does not carry over to quantum mechanics, because Heisenberg's uncertainty relation $\Delta x \Delta p \geq \hbar/2$ does not allow to specify a point in phase space with zero error. 
Thus workers have studied chaotic phenomena of a quantum system in a semi-classical regime, e.g. highly excited states of a hydron atom (Rydberg states). In this regime Gutzwiller's trace formula \cite{Gutzwiller:90}
derived from the path integral has been very useful. 
Also it has been quite popular to look for signatures of chaos in quantum systems, which have classically chaotic counter parts. According to Bohigas' conjecture \cite{Bohigas:84}, quantum chaos manifests itself in a 
characteristic spectral distribution, being of Wigner-type.
The spectral distribution is a global property of the system.

One may ask the question: Is it possible to obtain more specific and local information about the chaotic behavior of a quantum system? Like quantum mechanics for a metal conductor explains the existence conducting bands and non-conducting bands, there may co-exist regular and chaotic domains in a quantum system (as they are known to exist in
classical Hamiltonian chaos). 
To answer this question it is natural to look for a bridge between classical physics and quantum physics. 
The virtue of Gutzwiller's trace formula is that it forms such a bridge, valid in the semi-classical regime. 
More generally, a bridge between classical physics and quantum physics is renormalisation and the effective action. The effective potential $V^{eff}$ and effective action $\Gamma$ introduced in quantum field theory some decades ago \cite{Jona:64,Coleman:73,Dolan:74} is defined also in quantum
mechanics, viewed as QFT in $0+1$ dimensions.  
\bea
Z[J] &=& e^{-iW[J]}
\nonumber \\
\frac{ \partial }{ \partial J(x) } W[J] &=& - <0|\phi(x)|0>_{J}
\nonumber \\
\phi_{cl}(x) &=& <0|\phi(x)|0>_{J}
\nonumber \\
\Gamma[\phi_{cl}] &=& - W[J] - \int d^{4}y ~ J(y) \phi_{cl}(y) .
\eea
Cametti et al. \cite{Cametti:99} have computed the effective action $\Gamma$ in quantum mechanics, using perturbation theory and the loop expansion. 
They consider the Lagrangian
\be
L(q,\dot{q},t) = \frac{m}{2} \dot{q}^{2} - V(q), ~~~
V(q) = \frac{m}{2} \omega^{2} q^{2} + U(q) ,
\ee
where $U(q)$ is, e.g., a quartic potential $U(q) \sim q^{4}$.
Then the effective action takes the form
\begin{eqnarray}
\Gamma[q] &=& \int dt \left( - V^{eff}(q(t)) + \frac{Z(q(t))}{2} \dot{q}^{2}(t)
+ A(q(t)) \dot{q}^{4}(t) + B(q(t)) (d^{2}q/dt^{2})^{2}(t) + \cdots \right)
\nonumber \\
V^{eff} &=& \frac{1}{2}m \omega^{2} q^{2} +U(q) + \hbar V^{eff}_{1}(q) + O(\hbar^{2})
\nonumber \\
Z(q) &=& m + \hbar Z_{1}(q) + O(\hbar^{2})
\nonumber \\
A(q) &=& \hbar A_{1}(q) + O(\hbar^{2})
\nonumber \\
B(q) &=& \hbar B_{1}(q) + O(\hbar^{2}) .
\end{eqnarray}
There are higher loop corrections to the effective potential $V^{eff}$ as well as to the mass renormalisation $Z$. The most important property is the 
occurrence of higher time derivative terms. Actually, there is an infinite series of increasing order. This is due to a non-local structure coming from the expansion around a time-dependent classical path. Here comes the problem: When we want to interpret $\Gamma$ as effective action, e.g. for the purpose to analyze quantum chaos, the higher time derivatives require more boundary conditions than the classical action. 
Those boundary conditions are unknown. Moreover, analytical evaluation of the 
effective action to higher order in perturbation theory becomes prohibitively difficult and finally, such perturbation expansion does not converge. 
This makes the effective action in its standard form a tool unsuitable for
the analysis of quantum chaos.

\bigskip

\noindent {\bf 2. Quantum action} \\
In the following we will present an alternative way to construct an action taking into acount quantum corrections.
The classical trajectory from $x_{in}$ to 
$x_{fi}$ corresponds in quantum mechanics to the probability amplitude 
for the transition, given in terms of the path integral by
\be
\label{PathIntegral}
G(x_{fi},t_{fi};x_{in},t_{in}) = \left. \int [dx] \exp[ \frac{i}{\hbar} 
S[x] ] \right|_{x_{in},t_{in}}^{x_{fi},t_{fi}} .
\ee
The classical trajectory $x_{cl}$ is defined as extremum of the classical action
\be
\label{ClassAction}
S= \int dt L(x,\dot{x}) = 
\int dt \frac{m}{2} \dot{x}^{2} - V(x) .
\ee
In Ref.\cite{Jirari:99} we have proposed a conjecture, 
establishing a new link between 
classical mechanics and quantum mechanics. \\
{\em Conjecture:}
For a given classical action $S$ with a local interaction $V(x)$ 
there is a renormalized/quantum action 
\be
\label{QuantumAction}
\tilde{S} = \int dt \frac{\tilde{m}}{2} \dot{x}^{2} - \tilde{V}(x) ,
\ee
such that the transition amplitude is given by
\bea
\label{DefQuantumAction}
&& G(x_{fi},t_{fi}; x_{in},t_{in}) = \tilde{Z} 
\exp \left[ \frac{i}{\hbar} \left. \tilde{S}[\tilde{x}_{class}] 
\right|_{x_{in},t_{in}}^{x_{fi},t_{fi}} \right] ,
\eea
where $\tilde{x}_{class}$ denotes the classical path corresponding 
to the action $\tilde{S}$. 
The renormalized action is independent of the boundary points 
$x_{in}$, $x_{fi}$, but depends on the transition time $T= t_{fi}-t_{in}$.
Because the renormalized action, describing quantum physics, 
has mathematically the form of a classical action, 
this bridges the gap between classical and quantum physics. It 
opens a new vista to investigate quantum phenomena, the definition of which comes from classical physics. Prominent examples are instantons and chaos.

\bigskip

\noindent {\bf 3. Effective action vs. quantum action} \\
The effective action is defined via the vacuum-vacuum transition amplitude at infinite time. It allows to compute the ground state energy. 
The effective action is a function of the classical path. 
The quantum action is defined as transition amplitude from a set of initial positions to some set of final positions. It depends on some finite transition time. It does not depend on a classical path. 
In order to compare numerically the effective action with the quantum action, we have considered a harmonic oscillator with a weak anharmonic perturbation,
\be
\label{WeakAnhOscAction}
S = \int dt ~ \frac{m}{2} \dot{x}^{2} + v_{2} x^{2} + \lambda v_{4} x^{4}, ~~~ \hbar = m = v_{2} = m \omega^{2}/2 = v_{4} = 1.
\ee
We have varied the parameter $\lambda$ in the range $0 \leq \lambda \leq 0.1$ (weak perturbation). The effective action to one loop order and up to 2nd order of $\lambda$ yields \cite{Cametti:99}
\bea
v^{eff}_{2} &=& v_{2} + \delta v_{2}, ~~~ \delta v_{2} = \frac{3}{m \omega} \lambda 
\nonumber \\
v^{eff}_{4} &=& v_{4} + \delta v_{4}, ~~~ \delta v_{4} = \frac{9}{m \omega^{2}} \lambda^{2} .
\eea
The comparison with the quantum action, computed numerically for transition time $T=4$, is shown in Fig.[1].
One observes that the effective potential and the potential of the quantum action  are close for small $\lambda$. At about $\lambda \approx 0.06$, one observes a discrepancy, which shows the onset of non-perturbative effects.

We would like to point out that the quantum action in imaginary time can be interpreted as quantum action at finite temperature.
According to the laws of quantum mechanics and thermodymical equilibrium,
the expectation value of some observable $O$, like e.g. average energy
is given by
\be
<O> = \frac{ Tr\left[ O ~ \exp[ - \beta H] \right] }
{ Tr\left[ \exp[ - \beta H] \right] }
\ee
where $\beta$ is related to the temperature $\tau$ by 
$\beta = 1/(k_{B} {\tau})$.
On the other hand the (Euclidean) transition amplitude is given by
\begin{equation}
G(x_{fi},T;x_{in},0) = <x_{fi}| \exp[ - H T/\hbar ]|x_{in}>
\end{equation}
Thus from the definition of the quantum action, Eqs.(\ref{QuantumAction},\ref{DefQuantumAction}),
one obtains
\begin{equation}
<O> = \frac{ \int_{-\infty}^{+\infty} dx \int_{-\infty}^{+\infty} dy
<x|O|y> \exp[-\tilde{S}_{\beta}|_{x,0}^{y,\beta}] }
{ \int_{-\infty}^{+\infty} dx \exp[ - \tilde{S}_{\beta}|_{x,0}^{x,\beta}] } ,
\end{equation}
if we identify 
$\beta = \frac{1}{k_{B} {\tau}} = T/\hbar$.
As a result, the quantum action $\tilde{S}_{\beta}$ computed from transition time $T$, describes equilibrium thermodynamics at $\beta = T/\hbar$, i.e. temperature $\tau = 1/(k_{B} \beta)$.
Consequently, the quantum action for some finite transition time can be 
used for the study of quantum instantons and quantum chaos at 
finite temperature.

\bigskip

\vspace{0.2cm}
Tab.1. Double well potential $V(x) = \frac{1}{2} - x^{2} + \frac{1}{2} x^{4}$.  $T= 0.5$. \\
\begin{tabular}{|l|l|l|l|l|l|l|l|l|} \hline \hline
& $\tilde{m}$ & $\tilde{v_{0}}$ & $\tilde{v_{1}}$ & $\tilde{v_{2}}$ & $\tilde{v_{3}}$ & $\tilde{v_{4}}$  & \mbox{interval}  \\ \hline
Fit MC & 0.9959(1)  & 1.5701(54) & 0.000(2)  & -0.739(5)  & 0.000(2)  & 0.487(4)  &  [-1.2,+1.2] \\ \hline
Fit MC & 0.9961(2)  & 1.5714(17) & 0.000(2)  & -0.747(10) & 0.000(2)  & 0.489(7)  &  [-1.4,+1.4] \\ \hline
Fit MC & 0.9961(1)  & 1.5732(11) & 0.000(3)  & -0.760(5)  & 0.000(3)  & 0.499(3)  &  [-1.6,+1.6] \\ \hline
Fit MC & 0.9959(1)  & 1.5713(10) & 0.000(2)  & -0.747(4)  & 0.000(2)  & 0.495(2)  &  [-1.8,+1.8] \\ \hline
Fit MC & 0.9959(1)  & 1.5744(11) & 0.000(3)  & -0.754(4)  & 0.000(2)  & 0.498(2)  &  [-2.0,+2.0] \\ \hline
Fit MC & 0.9959(2)  & 1.5694(19) & 0.000(2)  & -0.739(6)  & 0.000(2)  & 0.492(3)  &  [-2.2,+2.2] \\ \hline
Fit MC & 0.9964(2)  & 1.5718(16) & 0.000(3)  & -0.745(6)  & 0.000(2)  & 0.491(2)  &  [-2.4,+2.4] \\ \hline
Fit MC & 0.9962(8)  & 1.5685(18) & 0.000(2)  & -0.740(7)  & 0.000(1)  & 0.492(3)  &  [-2.6,+2.6] \\ \hline
Fit MC & 0.9963(2)  & 1.5731(7)  & 0.000(0)  & -0.742(2)  & 0.000(2)  & 0.492(1)  &  [-2.8,+2.8] \\ \hline
Fit MC & 0.9966(3)  & 1.5695(2)  & 0.000(4)  & -0.744(8)  & 0.000(3)  & 0.492(2) &  [-3.0,+3.0] \\ \hline
Average &  0.9961(2) & 1.5710(17) & 0.000(2) & -0.745(6) & 0.000(2) & 0.493(3) &   \\ \hline \hline
\end{tabular}
\\

\bigskip

\noindent {\bf 4. Quantum Instantons} \\
Instantons are classical solutions of field theories in imaginary time 
(Euclidean field theory) \cite{Rajaraman:82}. 
Instantons are believed to play an important role in gauge theories. 
They are the tunneling solutions between the $\theta$-vacua \cite{Jackiw:76}.
t'Hooft has solved the $U_{A}(1)$ problem \cite{tHooft:76}, showing that 
due to instanton effects, the Goldstone boson is not a physical particle 
in this case. In $QCD$ instantons may be responsible for quark 
confinement. It has recently been predicted by Shuryak \cite{Shuryak:00} 
that $QCD$ at high temperature and density not only displays a transition 
to the quark-gluon plasma, but has a much richer phase structure, due to 
strong instanton effects. Instantons in gauge theories can be characterized
by topological quantum numbers. 
Tunneling and instantons play an important role also in the inflationary 
scenario 
of the early universe. During inflation, quantum fluctuations of the 
primordial field expand exponentially and terminate their evolution as a classical field. 
Its deviation from its average value is of the order of the size of the 
horizon \cite{Starobinsky:79}. The evolution of those clssical fluctuations 
eventually leads to the formation of galaxies \cite{Khlopov:98}. There are 
theoretical 
contradictions in scenarios where inflation terminates by a first order 
phase transition \cite{Kolb:91}.

Here we suggest that besides the classical instanton solution
also its quantum descendent can be precisely defined and  
quantitatively computed. We want to discuss this matter in the context of 
quantum mechanics. We consider in 1-D a particle of mass 
$m$ interacting with a quartic potential, which is symmetric under parity 
and displays two minima of equal depth (double well potential). This is an 
analogue of "degenerate vacua" occuring in field theory. The potential
\be
\label{PotDegVac}
V(x) = A^{2}(x^{2}-a^{2})^{2},\ee
has minima located at $x= \pm a$.
In particular, we have chosen the potential
$V(x) = \frac{1}{2} - x^{2} + \frac{1}{2} x^{4}$ ($A = 1/\sqrt{2}$, $a=1$),
and
$m = 1$, $\hbar = 1$.
The double well potential has a classical instanton solution. It is obtained by 
solving the Euler-Lagrange equations of motion of the Euclidean classical 
action, with the initial conditions $x(t=-\infty) = -a$, $\dot{x}(t=-\infty)=0$,
\bea
\label{ClassInstanton}
x^{cl}_{inst}(t) = a ~\mbox{tanh}[ \sqrt{2/m}~ A ~a ~t] .
\eea
The classical instanton goes from
$x(t=-\infty) = -a$ 
to $x(t=+\infty) = +a$ (see Fig.[3]).

What is the problem with the quantum instanton? 
Due to Heisenberg's uncertainty relation, there is no quantum solution corresponding to the initial conditions of the classical instanton 
(specifying initial position and momentum with zero uncertainty). The same problem is the reason why one can not define Lyapunov exponents in quantum chaos, and hence can not apply the concepts of classical chaos. 
In the following we will define the quantum instanton via the  
renormalized action, for which we make the following ansatz,
\be
\label{RenActionDegVac}
\tilde{S} = \int dt \frac{1}{2} \tilde{m} \dot{x}^{2} - \sum_{k=0}^{4} \tilde{v}_{k} x^{k}.
\ee
Note that one expects higher terms to occur in the renormalized potential. 
In a first step, we have restrained the search for the renormalized potential 
parameters to fourth order polynomials, for the sake of numerical simplicity. 
The term $\tilde{v}_0$ represents the constant term $\tilde{Z}$ in 
Eq.(\ref{DefQuantumAction}). We have worked in imaginary (Euclidean) time.
We have computed the transition matrix element $G$, 
Eq.(\ref{DefQuantumAction}), by using the Monte Carlo method and as alternative 
from the spectral decomposition by solving the Schr\"odinger equation for the lowest 7 and 30 states, respectively.

\bigskip

\noindent The parameters $\tilde{m}$ and $\tilde{v}_{k}$ have been obtained 
by minimizing the difference between the l.h.s. and the r.h.s. of 
Eq.(\ref{DefQuantumAction}), for a given time interval $T=t_{fi}-t_{in}$ and 
different combinations of boundary points $x_{fi}$, $x_{in}$. For more details 
about the numerical method see Ref.\cite{Jirari:99}.  
Results of the renormalized action at $T=0.5$ from Monte Carlo are presented in 
Tab.[1]. The renormalized parameters appear to be reasonably independent
from the position of boundary points, distributed over the interval $[-a,+a]$.
The data correspond to $J=6$ initial and final boundary points.

The renormalized action parameters as a function of $T$ are shown in Fig.[2a,b]. Fig.[2a] shows the mass $\tilde{m}$ and the constant term 
$\tilde{v}_{0}$ of the potential as a function of $T$, while Fig.[2b] shows the quadratic term $\tilde{v}_{2}$ and the quartic term $\tilde{v}_{4}$. 
The data in Fig.[2] can be interpreted also parameters of the quantum action 
versus finite temperature $\tau$ ($\tau = \hbar/(k_{B} T)$), and $T \to \infty$ corresponds to the zero temperature limit.
We have introduced the time scale $T_{sc} = \hbar/E_{gr}$, 
via the ground state energy $E_{gr}=0.568893$, giving $T_{sc}=1.75779$. 
Over a wide range of $T$ values, one observes that $\tilde{v}_{0}$ 
can be fitted by $\tilde{v}_{0} \sim A + B/T$. For $T \to \infty$ one 
finds that the asymptotic behavior $\tilde{v}_{0} \to 0.5677 \pm 0.015$, 
compatible with $\tilde{v}_{0} \to E_{gr}$. The 
mass $\tilde{m}$ changes little in the regime $0 < T < T_{sc}$,
it undergoes a drastic change between $T_{sc} < T < 7$, and 
stabilizes for $T > 7$. 
The upper part of error bars indicate the error of the fit, while the lower part of error bars indicate $\sigma$ of the estimator under variation of intervals 
of boundary points. It turns out that both errors are quite small for $\tilde{v}_{0}$. One observes asymptotic convergence of the renormalized parameters when $T/T_{sc}$ becomes large. On the other hand, the region of large $T$ is also the regime of the instanton (which goes from $t=-\infty$ to $t=+\infty$). Thus we suggest the following \\

\noindent {\em Definition of the quantum instanton:}
The quantum instanton is the classical instanton solution of the renormalized action in the regime of asymptotic convergence in time $T$ (zero temperature). 
The quantum instanton at finite temperature $\tau$ is the classical instanton solution of the renormalized action at the corresponding finite transition time $T$.  \\

\noindent For example, at $T=0.5$ we find the following parameters 
$\tilde{m}=0.9961(2)$,  
$\tilde{v}_{0}=1.5710(17)$, 
$\tilde{v}_{1}=0.000(2)$, 
$\tilde{v}_{2}=-0.745(6)$, 
$\tilde{v}_{3}=0.000(2)$, 
$\tilde{v}_{4}=0.493(3)$. 
By adding a constant, the renormalized potential can be expressed as
\be
\label{RenPotDegVac}
\tilde{V}(x) = \tilde{A}^{2}(x^{2}-\tilde{a}^{2})^{2},
\ee
where $\tilde{A}^{2} = \tilde{v}_{4}$ 
and $\tilde{a}^{2} = -\tilde{v}_{2} / (2 \tilde{v}_{4})$,
giving $\tilde{A} = 0.702(2)$ and $\tilde{a}=0.869(6)$.
The minima of the potential $\tilde{V}$ are located at $\pm \tilde{a}$.
Thus, like the classical action, also the renormalized action at $T=0.5$ displays "degenerate vacua". Hence it has an instanton solution,
corresponding to $T=0.5$, given by 
\be
x_{inst}^{T=0.5}(t) = \tilde{a} ~\mbox{tanh}[\sqrt{2/\tilde{m}} ~\tilde{A} ~\tilde{a} ~t] 
\approx 0.869 ~ \mbox{tanh} [0.865 ~ t ] .
\ee
Similarly, we find an instanton solution for any larger value of $T$. The 
quantum instanton is obtained in the asymptotic limit $T \to \infty$.
The evolution of the instantons under variation of $T$, i.e. under variation of the temperature, is depicted in Fig.[3]. It shows the transition from the 
classical instanton (at infinite temperature) to the quantum instanton
(at zero temperature).  
One observes asymptotic convergence for $T>7$ and a drastic change 
in between the classical instanton ($T=0$) and the quantum instanton 
($T\geq 9$).
The corresponding instanton action is shown in Fig.[4]. While the action of 
the classical instanton is of order $\hbar = 1$, the action of the quantum 
instanton is smaller by one order of magnitude. This is due to a lower barrier of the potential in the quantum action.

\bigskip

\noindent {\bf 5. New Vista to Quantum Chaos} \\
For classical chaos, we have today precise 
mathematical definitions of the concepts, many experimental observations 
and high precision computer simulations. 
However, the physical understanding of its quantum correspondence, 
the so-called quantum chaos \cite{Gutzwiller:90,Blumel:97} has not 
matured to the same degree. When we speak of quantum chaos we mean 
the quantum descendent of a system, which is classically chaotic. 
An example is the kicked rotor. In quantum mechanics it corresponds 
to an atom interacting with a pulsed laser. Workers have investigated chaos also in scattering systems \cite{Rankin:71}. The signals of quantum chaos are less clear in such situations.
For quantum systems one has neither been able 
to define nor to compute quantum Lyapunov exponents, 
quantum Poincar\'e sections,  
quantum Kolmogorov-Sinai entropy etc.
Let us consider as example a Hamiltonian system in D=2 dimensions, 
suggested by Henon and Heiles \cite{Henon:64} in the context 
of planetary motion. Its classical action is given by 
\be
\label{HenonHeilesAction}
S = \int dt \frac{1}{2}(\dot{x}^{2} + \dot{y}^{2})
- \frac{1}{2}(x^{2} + y^{2} + 2x^{2}y - \frac{2}{3}y^{3}) .
\ee
According to the conjecture, there is a renormalized action $\tilde{S}$
and all quantum transition amplitudes can be written as a sum over 
the classical paths corresponding to the renormalized action.
Then one can apply the tools of classical chaos theory 
to the action $\tilde{S}$ 
in order to decide if $\tilde{S}$ produces classical chaos.
On the other hand $\tilde{S}$ describes quantum physics. Thus we suggest 
the following criterion to decide whether the Henon-Heiles system is 
quantum mechanically chaotic: \\

\noindent {\em Definition of quantum chaos}: Consider a classical system, described by a classical action $S$. Its corresponding quantum mechanical system is said to display quantum chaos (at zeo temperature), if the renormalized action $\tilde{S}$ in the regime of asymptotic convergence in time $T$ displays classical chaos. Also we speak of quantum chaos at finite temperature $\tau$, if the renormalized action $\tilde{S}$ at the corresponding finite transition time $T$ is classically chaotic.

\bigskip

\noindent {\bf 6. Concluding Discussion} \\
What is the physical interpretation of the quantum action and its corresponding 
trajectory? First, in the path integral based on the classical action, quantum physics is due to a sum over histories (paths) including the classical path, but mostly paths different from the classical one.
In the quantum action, by construction, there is only one path, 
which is a clssical path of the quantum action.
Quantum physics effects are captured in the coefficients of the   
new action. What is the interpretation of the trajectory of the quantum action?
Is there a particle following this trajectory?
We have introduced an effective (quasi) particle.
Its behavior differs from that of a quantum particle. E.g., we have seen that 
it has a different mass. Moreover, its behavior between initial and final points of propagation is different from a quantum mechanical particle:
It follows a smooth trajectory, while a quantum particle on average performs a zig-zag motion, (fractal curve of Hausdorff dimension $d_H = 2$). 
However, like the quantum mechanical particle, the effective particle leaves from  and arrives at the same boundary points, with the same probability amplitude! The concept of effective particles or quasi particles is well known in physics, e.g., describing collective degrees of freedom in nuclei or Cooper pairs in the BCS theory of supraconductivity.

\bigskip

\noindent {\bf Acknowledgements} \\
H.K. and K.J.M.M. have been supported by NSERC Canada. 
X.Q.L. has been supported by NSF for Distinguished Young 
Scientists of China, by Guangdong Provincial NSF and by the Ministry of Education of China.
H.K. is grateful for discussions with L.S. Schulman.

\newpage
\begin{flushleft}
{\bf Figure Caption}
\end{flushleft}
\begin{description}
\item[{Fig.1}]
Quantum corrections in action parameters for a weakly anharmonic perturbed harmonic oscillator, Eq.(\ref{WeakAnhOscAction}). Comparison of perturbative prediction from effective action versus quantum action. 
\item[{Fig.2}]
Double well potential. (a) Parameters $\tilde{m}$, $\tilde{v}_0$, and (b) parameters $\tilde{v}_2$ and $\tilde{v}_4$ of quantum action versus transition time $T$ (inverse temperature $\beta$).
\item[{Fig.3}]
Quantum instanton solution of quantum action corresponding 
to transition time $T$ (inverse temperature $\beta$).
\item[{Fig.4}]
Value of quantum action of quantum instanton solution corresponding to transition time $T$ (inverse temperature $\beta$).
\end{description}

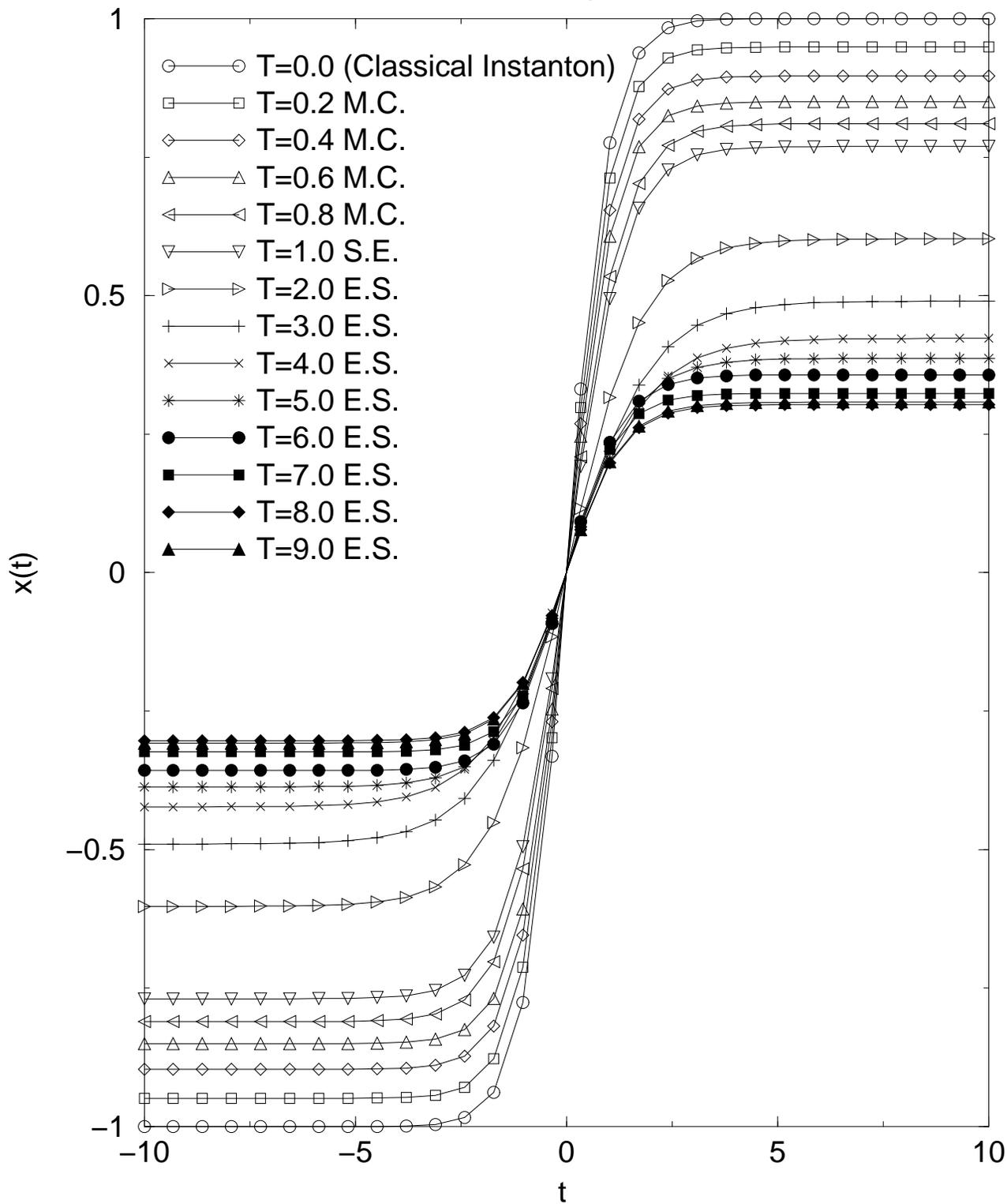
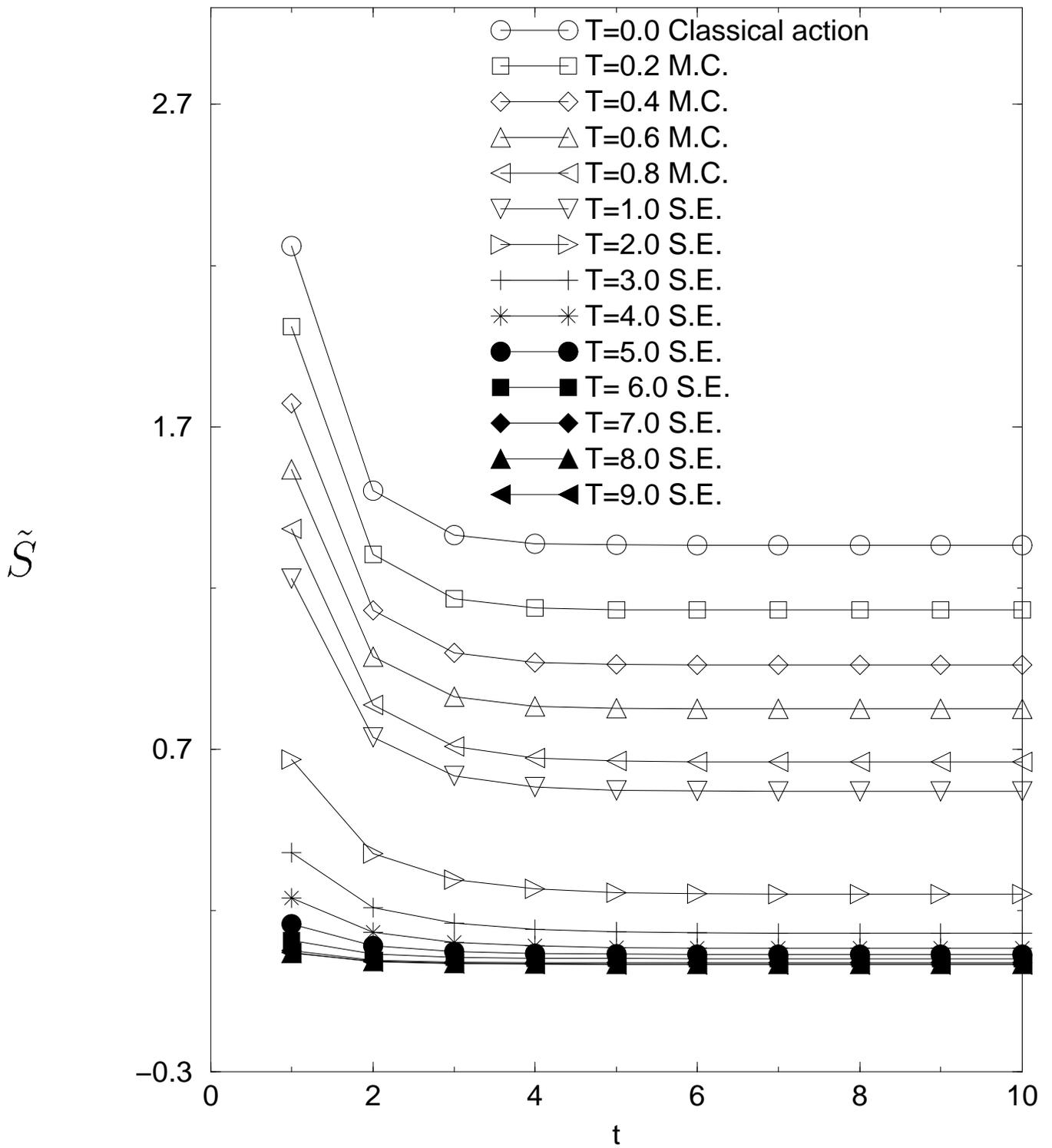
\end{document}